\documentclass[twocolumn,prx,superscriptaddress,
amsmath,amssymb]{revtex4-2}

\usepackage{graphicx}
\usepackage{dcolumn}
\usepackage{bm}
\usepackage{breqn}
\usepackage{floatrow}
\usepackage{braket,enumerate}
\usepackage{tikz}
\usetikzlibrary{calc,tikzmark}
\usepackage[hidelinks]{hyperref}

\usepackage{xcolor}
\definecolor{royalblue}{rgb}{.306,.451,.875}
\definecolor{firebrick}{rgb}{.698,.133,.133}
\definecolor{seagreen}{rgb}{.18,.545,.341}

\newcommand{\R}[0]{\mathbb R}
\newcommand{\C}[0]{\mathbb C}

\newfloatcommand{capbtabbox}{table}[][\FBwidth]

\allowdisplaybreaks[1]

\begin{document}

\title{Universal characterization of epitope immunodominance from a multi-scale
model of clonal competition in germinal centers}

\author{Federica Ferretti}
\affiliation{Department of Chemical Engineering\char`,{} Massachusetts Institute of Technology\char`,{} Cambridge\char`,{} Massachusetts 02139\char`,{} USA} 
\author{Mehran Kardar}
\affiliation{Department of Physics\char`,{} Massachusetts Institute of Technology\char`,{} Cambridge\char`,{} Massachusetts 02139\char`,{} USA}

\begin{abstract}
We introduce a novel, multi-scale model for affinity maturation, which aims to capture the intra-clonal, inter-clonal and epitope-specific organization of the B cell population in a germinal center. We describe the evolution of the B cell population via a quasispecies dynamics, with species corresponding to unique B cell receptors (BCRs), where the desired multi-scale structure is reflected on the mutational connectivity of the accessible BCR space, and on the statistical properties of its fitness landscape. Within this mathematical framework, we study the competition among classes of BCRs targeting different antigen epitopes, and construct an effective \emph{immunogenic space} where epitope immunodominance relations can be universally characterized. We finally study how varying the relative composition of a mixture of antigens with variable and conserved domains allows for a parametric exploration of this space, and identify general principles for the rational design of two-antigen cocktails.
\end{abstract}

\maketitle

\section{Introduction}

The molecular foundation of pathogen recognition and neutralization is the specific, high-affinity 
binding between antibodies and antigens \cite{Janeway}. Individual antibodies or B cell receptors recognize the antigen at discrete surface accessible regions, known as antigenic determinants or B-cell epitopes.
The size of these recognition sites is considerably smaller than the overall size of natural antigens; such as viral proteins or other pathogen-derived molecules. As a result, a population of antibodies interacting with the same antigen can give rise to a multitude of different structural conformations for the bound antigen-antibody complex.

While the accurate identification and prediction of B cell epitopes is still a laborious and expensive experimental task  and an outstanding computational challenge, a mixture of experimental and computational methods have been developed to extract a coarse-grained classification of groups of antibodies based on where they bind on the surface of a given antigen \cite{epitope-methods,science-SPR}. We can then see antigens, from the perspective of immune responders, as displaying a mosaic-like surface, with antibodies that bind on the same tile grouped into distinct \emph{classes}. The question we are interested in is when and why certain antibody classes can outnumber others, in the antibody repertoire of an individual or of a population.

The composition of each individual's antibody pool is the result of an accelerated evolutionary process called affinity maturation, occurring in the lymph nodes upon the encounter of a foreign agent. In response to this event, the immune system of the host organizes a sophisticated learning machinery in substructures of the lymph nodes known as germinal centers, where low-affinity naive B cells undergo iterated rounds of replication, mutation and selection to acquire the desired specificity. The resulting B cell population produced by this affinity maturation process will compose the memory and antibody repertoires of the infected host \cite{GCrev-2022}. 

Antigenic drift ---i.e. the immune evasion pattern of some viruses--- has revealed  that the primary antibody response is often focused towards a small subset of epitopes, referred to as \emph{immunodominant}, rather than being uniformly directed towards the entire set of possible target sites \cite{influenza_drift}. 
The physicochemical, structural, and geometrical aspects of the distinct antigen epitopes, as well as potential biases in the naive B cell repertoires, indeed cause the antigen surface to be non homogeneously immunogenic \cite{gerline-classification,Assaf-2018}. While in principle these differences in epitope immunogenicity could vary from one individual to another because of the personalized aspects of naive repertoires \cite{AT-review},  epitope immunodominance seems to be largely a property of the pathogen, only slightly affected by individual or even organismal particularities (at least for certain viruses) \cite{Altman-lamprey}. 

The key idea explored in this paper is that B cell epitope immunodominance can be described as a phenomenon emerging from the general statistical features of the affinity landscapes of competing antibody classes. 
We introduce a new minimal model
for affinity maturation, inspired by a long tradition of computational models, such as~
\cite{Oprea-Perelson, Shenshen, Marco_AM, Leerang-Matt, Kayla-cocktail, Leerang-chimera}, whose central feature is a multi-scale representation of the clonal competition in a germinal center. B cell competition indeed occurs at several different stages:
\begin{enumerate}[(i)]
    \item inter-class level: following the nomenclature used for SARS-CoV-2 antibodies, B cell receptors (BCRs) are categorized into \emph{classes} based on the epitopes that they bind to \cite{barnes-class}. 
    \item intra-class level: within each class, competition occurs among clonal lineages originating from different B cell ancestors (germlines); 
    \item intra-clonal level: somatic hypermutations occurring during affinity maturation produce variability in the fitness of B cells within the same clonal family, enabling competition even at this level.
\end{enumerate}
Within this framework of evolving B cell populations, more immunodominant epitopes are associated with higher fixation probabilities of the corresponding BCR classes.
We compute the fixation probability as an extreme value problem, and identify, for a specific class of fitness landscape models, a restricted set of universal parameter combinations which modulate immunodominance. 

The proposed model can be applied to immunodominance phenomena across different affinity maturation contexts.
For example, we  investigate how the epitope immunodominance relations can be manipulated by exposing the system to a cocktail of two antigens with varying relative concentrations. 
This is relevant to the problem of
vaccination by a cocktail of antigens, a procedure employed in the development of  vaccines against highly mutating pathogens, such as the SARS-CoV-2 bivalent booster \cite{bivalent-moderna, bivalent-mice}. The rationale behind the use of immunogen cocktails is that  simultaneous exposure to two or more antigen variants confers a competitive advantage to  \textit{broadly neutralizing antibodies}, i.e. antibodies capable of neutralizing multiple variants of a virus by recognizing conserved regions of the antigen. We employ our model in a toy description of the cocktail context to determine under what general conditions the manipulation of vaccine compositions can alter the immunodominance relations between conserved and variable epitopes.

A more detailed mathematical description of the dynamical model is provided in Sec.~\ref{sec:model}. Results for the construction of a universal immunodominance phase diagram and for the problem of optimizing a bivalent antigen cocktail are presented in Sec.~\ref{sec:results}; implications for future work and data analysis are briefly discussed in Sec.~\ref{sec:conclusion}.

\section{\label{sec:model}Model}
\subsubsection{A schematic description of affinity maturation}

Affinity maturation (AM) is the key training process enabling the mammal adaptive immune system to learn upon the encounter of a foreign agent. It encompasses an intricate series of reactions, involving various lymphocytes, cytokines and signaling pathways \cite{GCrev-2012,GCrev-2022,GCrev-dyn}. At a coarse-grained level, AM is akin to an iterative two-step evolutionary process that takes place in germinal centers (GCs) formed in lymph nodes upon encounter on pathogen: Some (germline) B cells with some affinity to the antigen first undergo replication with a fast accumulation of mutations (in the dark zone of GC), followed by competitive selection (in the light zone of GC). Through repetition of these two steps, the B cell repertoire of an individual is expanded and refined in affinity for the encountered antigen.

We adopt here a simplified model for affinity maturation, which describes the evolution of B cells via parallel mutation, replication and death events. Competition is introduced by looking at the dynamics of the \textit{fraction} of each BCR sequence in the germinal center, rather than their absolute number, as in Eigen's well-known quasispecies model \cite{Eigen-book}. In GCs all  B cells compete for the same resources, i.e. interactions with follicular T cells and antigen capture.
We consider each productive B cell receptor sequence as a distinct quasispecies, denoted by index $i=1,\dots,N$. 
Replication, death and mutation events are described as simple first order reactions:
\begin{equation}
    B_i\xrightarrow{\lambda_i} B_i+B_i,\quad B_i\xrightarrow{\delta} \emptyset,\quad B_i\xrightarrow{\mu_{ji}} B_j\,,
    \label{reactions}
\end{equation}
where $\lambda_i$ is a replication rate that depends on the affinity of the BCR to the presented antigen, $\delta$ is a constant death rate, and $\mu_{ji}$ is the effective mutation rate from sequence $i$ to sequence $j$. For the sake of simplicity, we assume that mutations only occur at a constant rate $\mu$ between pairs of productive sequences at a Hamming distance of one \footnote{This simplifying assumption overlooks the current understanding of the statistics of hypermutations, which have been observed to be far from uniform and 
to result into enrichment of specific nucleotide motifs \cite{SHM-CSR,DiNoia,Nati-SHM,AT-review}. 
Nonetheless, we believe that a more realistic depiction of the mutational network will not significantly change our qualitative results and is not considered in this work.}.

\subsubsection{Structure of the B cell receptor space}
Once the hopping rate is fixed, only the connectivity of the graph of allowed mutations needs to be specified. In accord with the choice of a homogeneous rate, we decide to focus here only on the \emph{likely} mutations which can accumulate on any BCR sequence during the process of affinity maturation \cite{Nati-SHM}. For each clonal lineage, originating from a distinct germline ancestor, let us encode in a binary string of length $d$ any of such accessible sequences, so that the resulting mutational graph associated to the lineage is a $d$-dimensional hypercube. 

Since multiple germlines are recruited to initiate affinity maturation, we describe the germinal center as a collection of disconnected hypercubes. The mutual distance between two distinct germlines is indeed typically larger than the mutational distance between somatic B cells and their germline ancestors (usually of the order of 10 residues or less \cite{memory-mutations,Cell-torso}), suggesting that events of convergent evolution can be generally neglected.

To capture the competitive dynamics of B cell sub-populations targeting distinct parts of the antigen, we finally group the clonal lineages into classes \footnote{We note that the class definition is somewhat arbitrary: for an example of experimental methods used to identify antibody epitope classes, see \cite{barnes-class} or \cite{science-SPR}}. To each class we assign a specific distribution of fitness landscapes, so that the landscapes attributed to all lineages within the same class are treated as independent, quenched realizations derived from these distributions. 
This model is based on the premise that the fitness of a B cell receptor is largely determined by its binding affinity to the presented antigen, and that the specific geometry and chemical properties of each epitope will sculpt affinity landscapes sharing similar statistical features for groups of B cells within the same class. Our goal is to describe immunodominance in terms of the statistical differences among these class-specific fitness landscape ensembles.

\begin{figure*}
    \begin{minipage}{0.24\textwidth}
        \includegraphics[width=.95\textwidth,trim={1cm 0.1cm 0 0},clip]{  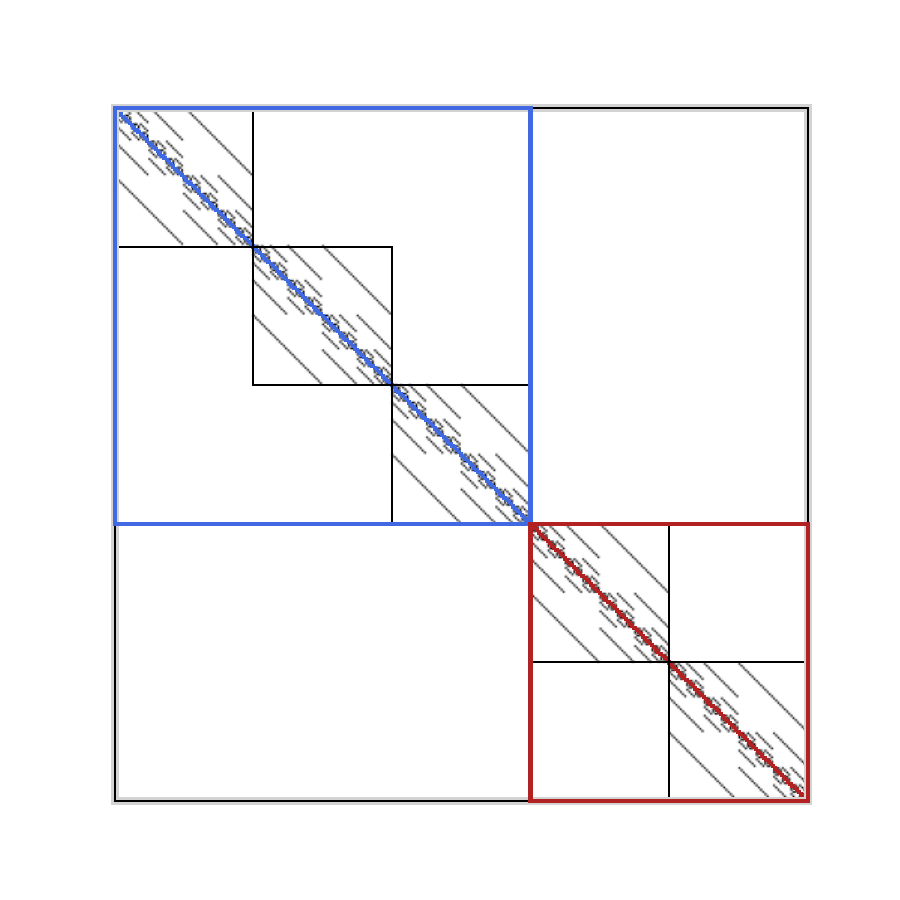}\llap{\parbox[b]{1.7in}{\textbf{Germinal Center:}\\\rule{0ex}{2.1in}}}\llap{\parbox[b]{1.65in}{\text{$\partial_t \bold n(t) =-\hat H\bold n(t)$}\\\rule{0ex}{1.85in}}}\llap{\parbox[b]{1.68in}{\text{$\hat H\in \R^{2^dM\times 2^dM}$}\\\rule{0ex}{1.55in}}}
    \end{minipage}
    \hfill
    \begin{minipage}{0.75\textwidth}
        \includegraphics[width=\textwidth,trim={3cm 4cm 4cm 4cm},clip]{  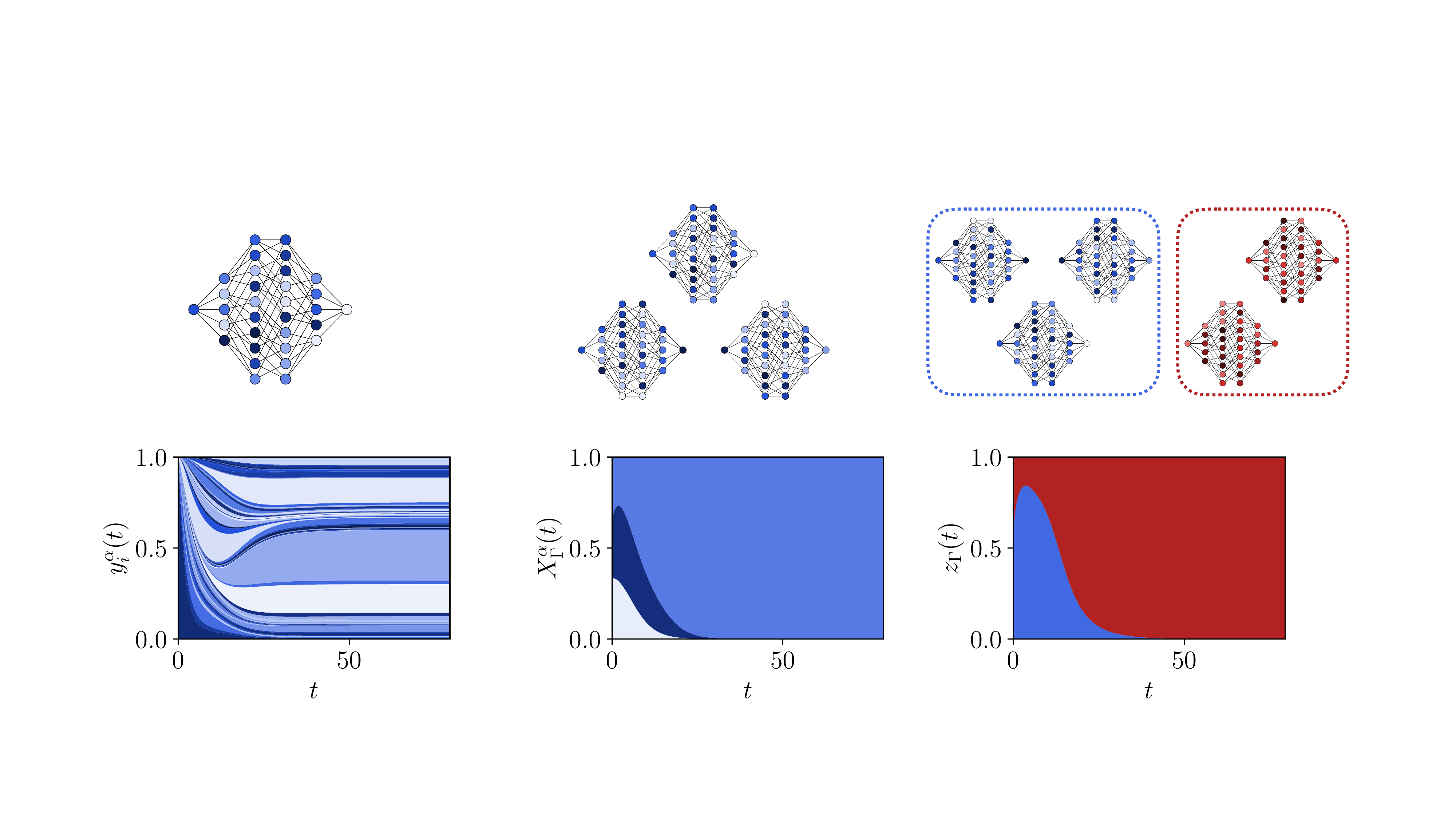}\llap{\parbox[b]{9.2in}{\textbf{1. Intra-clonal}\\\rule{0ex}{2.3in}}}\llap{\parbox[b]{5.8in}{\textbf{2. Intra-class}\\\rule{0ex}{2.3in}}}\llap{\parbox[b]{2.3in}{\textbf{3. Inter-class}\\\rule{0ex}{2.3in}}}
    \end{minipage}
    \begin{tikzpicture}[overlay,remember picture]
    \draw[arrows=-,black, semithick]  ( $ (pic cs:a) +(-47mm,4mm) $ ) --   ( $ (pic cs:a) +(-47mm,62mm) $ );
    \end{tikzpicture}
    \caption{Structure of the affinity maturation model. B cell evolution in a germinal center is described by a parabolic Anderson model (PAM) on a collection of disconnected $d$-dimensional hypercubes. Each clonal lineage evolves on a distinct hypercube, and lineages are grouped into classes, depending on the targeted antigen epitope. The resulting Anderson matrix has a block diagonal structure, with smaller blocks representing individual lineages/hypercubes. The diagonal entries are random variables drawn from the fitness distribution of the class corresponding to the outer, epitope-specific block. B cells are subject to an all-to-all competition, which manifests at three different levels (cf. Eqs.~\eqref{intra-clone}--\eqref{inter-class}). On the right panel, we show Muller plots for the evolution of quasi-species at each of these three levels, for a quenched realization of a germinal center seeded by $M=5$ germlines belonging to 2 distinct classes ($m_V=3, m_C = 2$), with Gaussian fitness distributions $p_V(f)=\mathcal N(1.6,0.5)$, $p_C(f)=\mathcal N(1.4,0.7)$.}
    \label{fig:hyper-scheme}
\end{figure*}
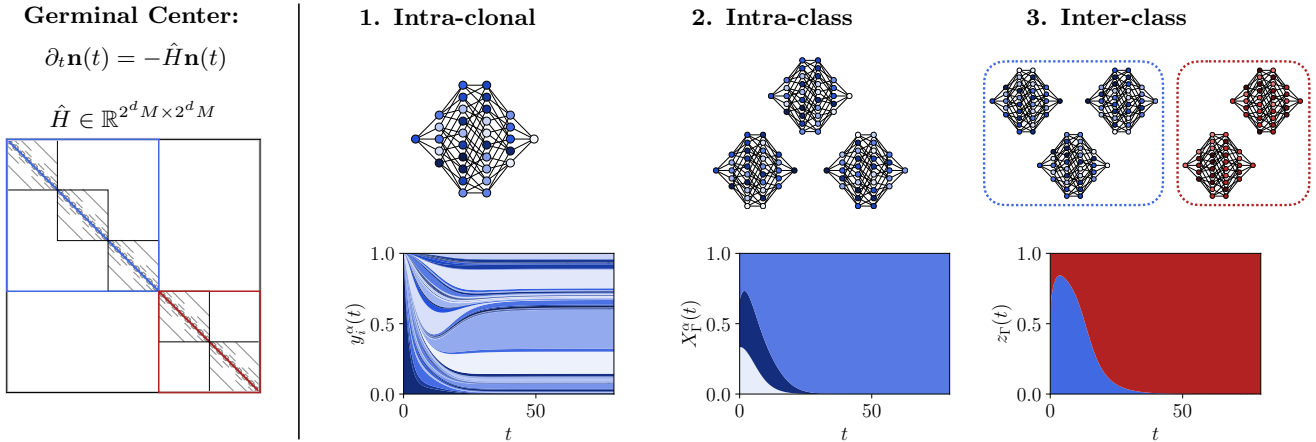

Figure~\ref{fig:hyper-scheme} provides an illustration of the resulting structure of the GC space. 

\subsubsection{Evolution dynamics} 

Let $n_i^\alpha$ be the expected size of quasispecies $i$ belonging to lineage $\alpha$; from \eqref{reactions}, we can derive a simple linear ODE for $\bold n(t)\in \R^{N\times M}$ :
\begin{equation}
    \partial_t n_i^\alpha = \left[f_i^\alpha \delta_{ij} + \mu\Lambda_{ij}\right]n_j^\alpha, \quad i=1,\dots,N,\ \alpha=1,\dots, M,
    \label{PAM}
\end{equation}
where $N=2^d$ is the number of nodes of each hypercube~\footnote{We make here the simplifying assumption that $d$ is constant.}, and $M$ is the number of clonal lineages in a single germinal center, typically of the order of $10-10^2$~\cite{GCrev-2022}. The $M$ clonal lineages are organized into a small number of classes, indexed by $\Gamma$, each of which contains $m_\Gamma$ elements. 
The $\Lambda_{ij}$ matrix indicates the Laplacian of the $d$-dimensional hypercube; the fitness is the net growth rate $f_i^\alpha = \lambda_i^\alpha -\delta^\alpha$. Initially the population is localized on the precursors' sequences: hence, without loss of generality, $n_i^\alpha(0)=\delta_{i,0}\ \forall \alpha$. 

In the case of independent identically distributed (IID) fitness variables, the model in Eq.~\eqref{PAM} is known as parabolic Anderson model (PAM) \cite{PAM-book}; analytical results for the PAM dynamics in the thermodynamic limit are known for several types of graphs, including hypercubes \cite{PAM-hypercube}.

We are interested here in the evolution of B cell species \emph{frequencies}. This is a common way to introduce competition in evolutionary models with only first order reactions, which has been demonstrated to capture the behavior of classical fixed population models (such as Wright-Fisher's or Moran's) 
in the limit of infinite population size \cite{Eigen-book}. 
Given the structure of the  model, we can focus on competition at different levels:
\begin{enumerate}[(i)]
    \item intra-clonal level:
    \begin{equation}
        \partial_t y_i^\alpha = \left[f_i^\alpha - \bar f^\alpha(t)\right]y_{i}^\alpha + \mu \sum_{j=1}^N\Lambda_{ij}y_j^\alpha,
        \label{intra-clone}
    \end{equation}
    where $y_i^\alpha(t)=n_i^\alpha(t)/\sum_{i=1}^N n_i^\alpha(t)$ is the fraction of identical clones $i$ within the lineage, and $\bar f^\alpha(t) = \sum_{i=1}^N f_i^\alpha y_i^\alpha(t)$ is the average fitness of the clonal family;
    \item intra-class level:
    \begin{equation}
        \partial_t X^\alpha_\Gamma = \left[\bar f^\alpha(t) - \Phi_\Gamma (t)\right]X_\Gamma^\alpha,
        \label{inter-clone}
    \end{equation}
    where $X^\alpha_\Gamma(t)=\sum_in_i^\alpha(t)/\sum_{\alpha\in\Gamma}\sum_in_i^\alpha(t)$ is the fraction of lineages in the class, and $\Phi_\Gamma(t) = \sum_{\alpha\in\Gamma} \bar f^\alpha (t)X^\alpha_\Gamma(t)$ is the population-averaged fitness of class $\Gamma$;
    \item inter-class level:
    \begin{equation}
        \partial_t z_\Gamma = \left[\Phi_\Gamma(t) - \bar F(t)\right]z_{\Gamma},
        \label{inter-class}
    \end{equation}
    where $z_\Gamma(t)= \sum_{\alpha\in\Gamma}\sum_{i}n_i^\alpha(t)/\sum_\alpha\sum_i n_i^\alpha(t)$ is the relative size of class $\Gamma$, and $\bar F(t)=\sum_\Gamma \Phi_\Gamma(t) z_\Gamma(t) $ is the total population-averaged fitness.
\end{enumerate}

At the intra-clonal level, the system reaches at long times a state of mutation-selection balance. 
At the intra-class and inter-class level, due to the disconnected structure of the global graph, the stable fixed points of Eqs.~\eqref{inter-clone}--\eqref{inter-class} correspond to fixation of the asymptotically fittest quasi-species and extinction of the rest.
The eventual dominance of a single clonal lineage in the GC population is consistent with experimental observations, even though affinity maturation typically terminates before the GC becomes completely monoclonal \cite{Victora-science, oligoclonal-flu}.
In this work, we will use the quenched average of $z_{\Gamma}(t)$ at long times to study epitope immunodominance from the GC response. 

In the asymptotic-time limit, for any class $\Gamma$, $z_\Gamma$ can only take values 0 or 1, depending on the realized collection of random fitness landscapes that describes the germinal center. The quenched average $\mathbb E[\lim_{t\to\infty}z_{\Gamma^*}(t)]$ is then equal to the fixation probability of class $\Gamma^*$, which can be computed as an extreme value problem:
\begin{widetext}
\begin{equation}
    \mathbb E [\lim_{t\to\infty} z_{\Gamma^*}(t)]=P_{fix,\Gamma^*}=m_{\Gamma^*}\int dx \rho_{\Gamma^*}(x)P_{\Gamma^*}(x)^{m_{\Gamma^*}-1}\prod_{\Gamma\neq\Gamma^*}P_{\Gamma}(x)^{m_{\Gamma}}\,,
    \label{Pfix}
\end{equation}
\end{widetext} 
where $\rho_{\Gamma}(x)$ is the probability density function (p.d.f.) of the asymptotic growth rate of the hypercube ``mass," and $P_{\Gamma}(x)$ is its cumulative density function (c.d.f.). The mass of hypercube $\alpha$ is defined as $N^\alpha(t)=\sum_{i}n_i^\alpha(t)$ and its asymptotic growth rate is 
\begin{equation}
    x^\alpha = \lim_{t\to\infty}\partial_t \log N^{\alpha}(t).
\end{equation}
Given the linear nature of Eq.~\eqref{PAM}, the asymptotic growth rate of any hypercube mass is given by the ground state eigenvalue of the matrix $H_{ij}^\alpha = -f_i^\alpha\delta_{ij} - \mu\Lambda_{ij}$ (corresponding to an Anderson Hamiltonian when the $f_i$'s are independent identically distributed, or IID, variables).
There is no general expression for the p.d.f. of the ground state eigenvalue of this random matrix, but we can derive approximate expressions using first order perturbation theory in two limiting regimes. 

Without loss of generality, let us fix the mutation rate to $\mu=1/d$: all growth rates $f_i^\alpha$ are then measured in rescaled units, such that one mutation per unit time is expected.
Let $\sigma_\Gamma$ denote the spread of the fitness values for any class $\Gamma$: the two limiting regimes are obtained when $d\sigma_\Gamma\ll 1$ (delocalized limit) or $d\sigma_\Gamma\gg 1$ (localized limit). The corresponding p.d.f.s for the ground state eigenvalues are (see App.~\ref{app:A}):
\begin{flalign}
    \label{rho-deloc}\rho^{del}_{\Gamma}(x) =& \int d\bold f\pi_\Gamma(\bold f)\,\delta\left(x-\frac{1}{N}\sum_{i=1}^N f_i\right);\\
    \label{rho-loc}\rho^{loc}_{\Gamma}(x) =& \int d\bold f \pi_\Gamma(\bold f)\, \delta\left(x - \max_{i=1\dots N }\{f_i\}+1\right);
\end{flalign}
where $\pi_\Gamma(\bold f)$ is the distribution of the realized disordered fitness landscape $\bold f$. When $d\sigma_\Gamma\ll1$, the asymptotic growth rate can be identified with the average fitness on the hypercube, thanks to the delocalized nature of the ground state eigenvector. This regime corresponds to a scenario in which competition within the same lineage is not strong. By contrast, a strong selection-weak mutation regime corresponds to the localized limit, $d\sigma_\Gamma\gg1$, where, neglecting the constant offset, the asymptotic growth rate can be identified with the extremum of the hypercube fitness values, where the ground state eigenvector is localized. 

\subsubsection{Antigen cocktails}\label{sec:cocktail}

\begin{figure*}[t]

\begin{floatrow}
\hspace{0.1in}

\begin{tabular}{c | ccc} 
  & $A$ & $B$ & $C$ \\ \hline\hline

  Antigen 1&  \checkmark & - & \checkmark \\ 
  Antigen 2 & - & \checkmark & \checkmark \\\hline
  \end{tabular}

\hspace{2.3in}

\textbf{Effective fitness landscapes}
\hspace{1.55in}
\end{floatrow}
\includegraphics[width=0.175\textwidth,trim = {0 3.5cm 0 0 },clip]{  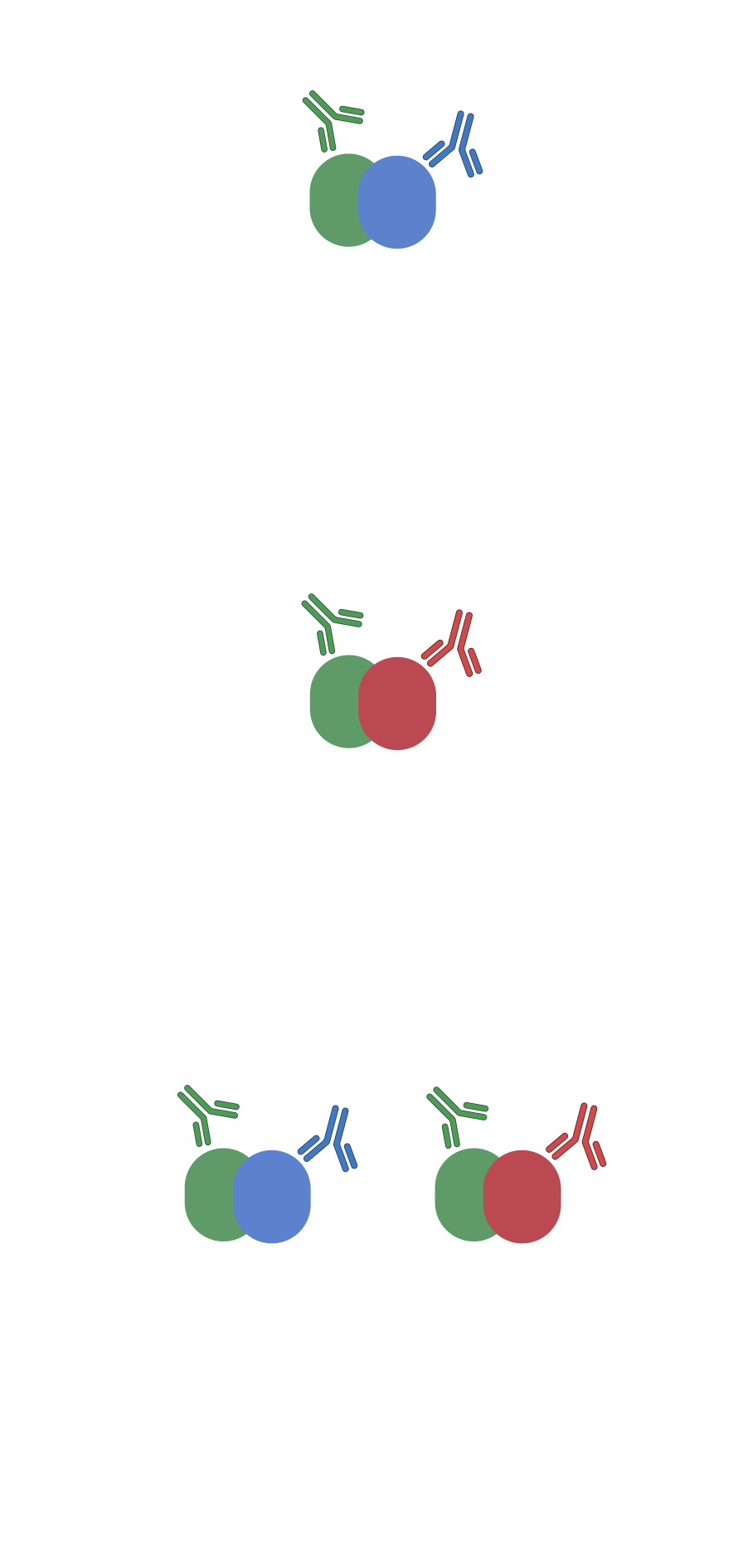}\llap{\parbox[b]{0.8in}{\textcolor{royalblue}{$A$}\\\rule{0ex}{2.15in}}} \llap{\parbox[b]{0.85in}{\textcolor{firebrick}{$B$}\\\rule{0ex}{1.32in}}} \llap{\parbox[b]{1.75in}{\textcolor{seagreen}{$C$}\\\rule{0ex}{2.15in}}} \llap{\parbox[b]{1.8in}{\textcolor{seagreen}{$C$}\\\rule{0ex}{1.32in}}} \llap{\parbox[b]{0.6in}{\text{Ag 1}\\\rule{0ex}{1.9in}}} \llap{\parbox[b]{0.7in}{\text{Ag 2}\\\rule{0ex}{1.15in}}} \llap{\parbox[b]{1.7in}{\text{$c$Ag 1 + $(1-c)$Ag 2}\\\rule{0ex}{0.57in}}} 
\hspace{0.1in}\hfill
\includegraphics[width=0.76\textwidth]{  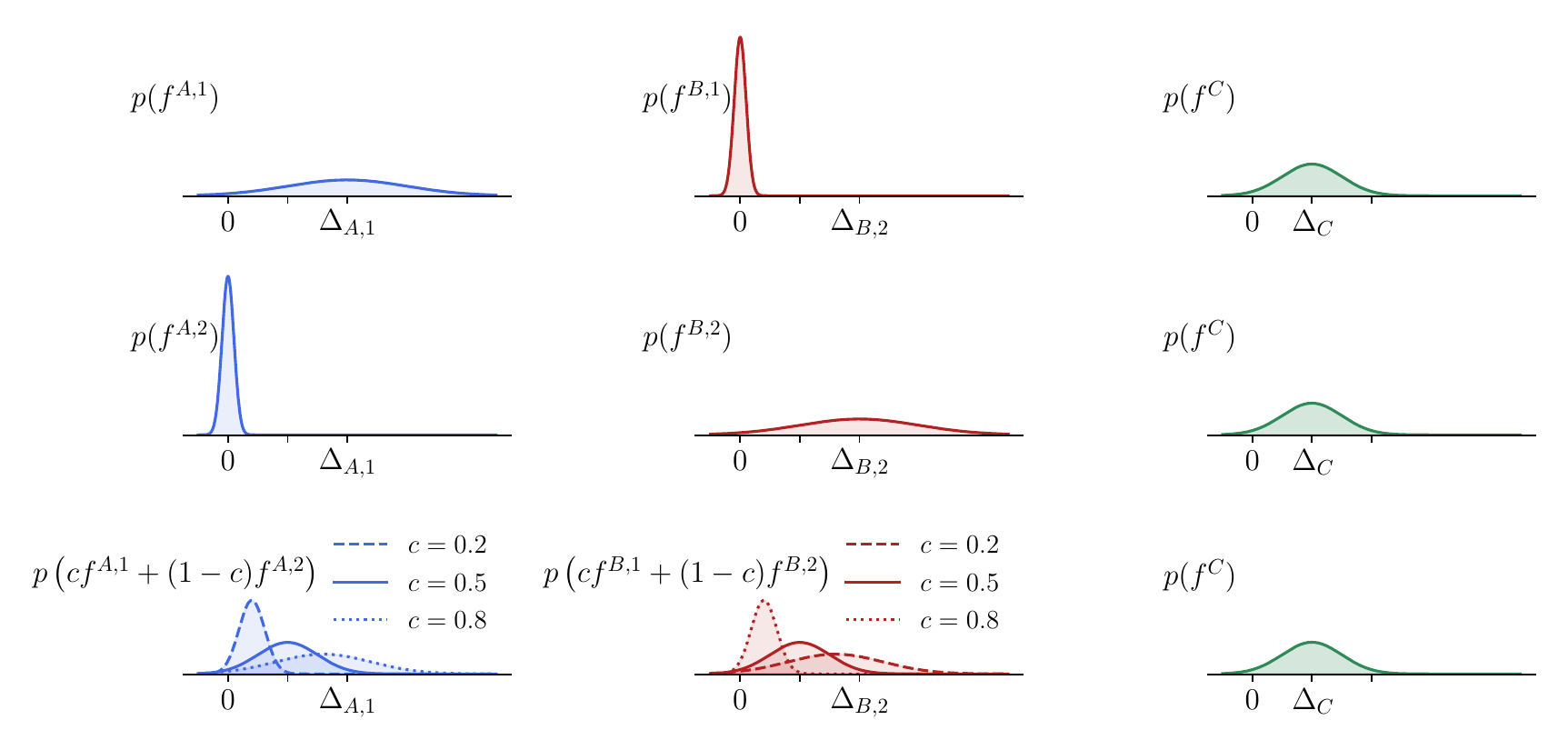}\llap{\parbox[b]{8.35in}{class $A$\\\rule{0ex}{2.5in}}}\llap{\parbox[b]{4.75in}{class $B$\\\rule{0ex}{2.5in}}}\llap{\parbox[b]{1.25in}{class $C$\\\rule{0ex}{2.5in}}}

\begin{tikzpicture}[overlay,remember picture]
    \draw[arrows=->,lightgray,ultra thick]  ( $ (pic cs:a) +(-55mm,33mm) $ ) --   ( $ (pic cs:a) +(-48mm,33mm) $ );
    \draw[arrows=->,lightgray,ultra thick]  ( $ (pic cs:a) +(-55mm,53mm) $ ) --   ( $ (pic cs:a) +(-48mm,53mm) $ );
    \draw[arrows=->,lightgray,ultra thick]  ( $ (pic cs:a) +(-55mm,13mm) $ ) --   ( $ (pic cs:a) +(-48mm,13mm) $ );
\end{tikzpicture}

\caption{The table summarizes the reactivity pattern of antigens and BCR classes. More precisely, reactivity is modeled by a Gaussian fitness distribution assigned to each antigen-BCR class pair. Non-reactive pairs have a narrow fitness distribution centered around zero, while reactive pairs have mostly positive fitness values. The effective “cocktail fitness landscape" is the weighted average of the fitness landscapes imposed by individual antigens, with weights equal to the relative antigen concentrations.  
The distribution of class $C$ remains unaffected by $c$ since the targeted epitope (in green) is the same on both antigens. }
\label{fig:cock-scheme}
\end{figure*}
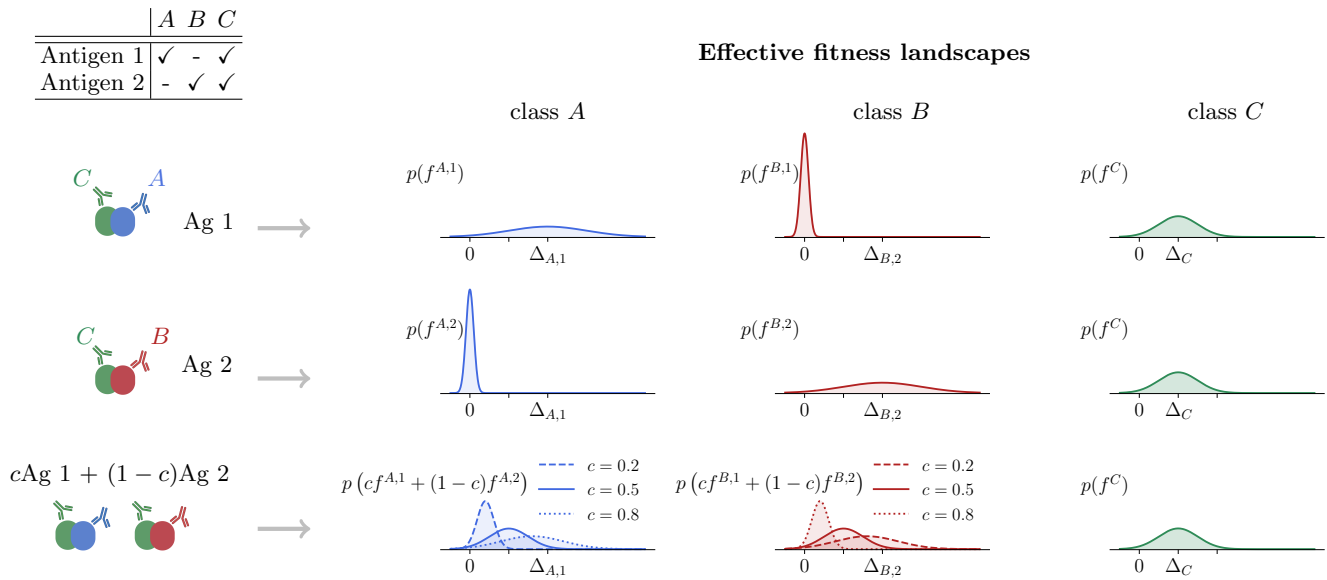

To model affinity maturation in the context of vaccination or infection by an antigen cocktail, we must specify the fitness landscape imposed by individual antigens on any class of BCRs, as well as the rules by which these landscapes are combined when the antigens are mixed. 

Let us focus on a simplified scenario, where only two antigens are included in the vaccine, in relative proportions $c$ and $1-c$. We further assume that the antigens are two variants of the same protein where we can distinguish, at a coarse-grained level, a mutated dominant epitope and a conserved subdominant epitope. The mutants are significantly distant in antigenic space, so that most of the antibodies generated after a primary immunization with one antigen may not neutralize the unseen antigen, as observed in the case of the SARS-CoV2 wild type strain and omicron variant \cite{sars-cov2-antigenic,bloom-escape}. Hence, at a first approximation, cross-reactivity can only be achieved by targeting the conserved subdominant epitope.

In this scenario, at least three classes of antibodies must be introduced: the classes targeting the mutant epitopes on antigen 1 and antigen 2, respectively named $A$ and $B$, and the class targeting the shared epitope ($C$). Let $f^{\Gamma,a}$ indicate the fitness value of a BCR belonging to class $\Gamma \in \{A,B,C\}$ under exposure to antigen $a\in\{1,2\}$. 
We can interpret $f_i^{\Gamma,a}dt$ as the probability to have a replication event for the BCR $i$ in a time interval $dt$, conditioned to the encounter of antigen $a$. Then the marginal probability of a replication event per unit time, in the presence of multiple antigens, reads:
\begin{eqnarray}
    f^{\Gamma}_i = c f^{\Gamma,1}_i + (1-c)f^{\Gamma,2}_i. 
    \label{f-cocktail}
\end{eqnarray}
This weighted average describes the effective fitness of the BCRs in the cocktail when all antigen types are presented homogeneously and abundantly on the follicular dendritic cells, so that each B cell, during its residency in the light zone, effectively samples their relative concentration. In the rest of this paper, we will work in this condition.

Similarly, the frequencies of activated precursors will be impacted by the composition of the cocktail. Let us assume that, for a fixed total amount of antigen, the expected number of precursors entering the GC is fixed to $M$, but the number of precursors of each class $m_\Gamma$ depends on the original abundance $\nu_\Gamma$ of reactive B cells of class $\Gamma$ in the repertoire and on the concentration of the targeted epitope. Imposing the condition $\sum_\Gamma m_\Gamma=M$, we have:
\begin{equation}
    \frac{m_A}{M} = \frac{c\nu_A}{Z(c)};\quad \frac{m_B}{M} = \frac{(1-c)\nu_B}{Z(c)};\quad \frac{m_C}{M} = \frac{\nu_C}{Z(c)};
    \label{germline-comp}
\end{equation}
where $Z(c) = c\nu_A + (1-c)\nu_B + \nu_C$.

 For the sake of simplicity, let us assume that the fitness associated to each non-reactive antigen-BCR pair is exactly zero, while fitness values associated to reactive antigen-BCR pairs are mostly positive. In that case, 
 \begin{equation}
     f_i^{A}  = c f_i^{A,1},\quad f_i^{B}  = (1-c) f_i^{B,2},\quad f_i^{C}  = f_i^{C},
     \label{fit-comp}
 \end{equation}
 where we assume that the fitness of the BCRs that target the unchanged epitope (class $C$), is unaffected by the antigen background. The same prescription as Eq.~\eqref{fit-comp} also applies to the effective asymptotic growth rates of the hypercubes' mass. Therefore:
\begin{widetext}
\begin{equation}
    P_{fix,C}=\int dx \frac{\partial}{\partial x}P_C(x)^{m_C(c)}P_A\left(\frac{x}{c}\right)^{m_A(c)}P_B\left(\frac{x}{1-c}\right)^{m_B(c)}.
\end{equation}
\end{widetext}

\section{The random energy model}\label{sec:results}

\subsubsection{Universality of immunodominance for REM landscapes}\label{sec:imm-space}

So far we described a flexible paradigm which can be adapted to any fitness landscape model. We must now specify what type of fitness landscape is associated to the hypercubes where the clonal lineages evolve.
Several theoretical models have been proposed over the decades to effectively model the topography of real fitness landscapes \cite{Krug-nature,HoC-Kingman,Franz-Peliti-REM,NK-Kaufman,mt-Fuji} and, since the advent of high-throughput sequencing techniques, an increasing number of them has been empirically reconstructed \cite{rev-biophys-landscape}.

In the case of affinity maturation, the fitness landscape is shaped by the presented antigen. In order to achieve reproducible antibody evolution, at least at the phenotypic level, the fitness of BCRs must be largely determined by their binding affinity to the antigen; a phenotype which has been long believed to have an almost-linear relation to the genotype \cite{Milstein-rev,GCrev-dyn}. However, recent experimental results have demonstrated that other random effects can contribute to determining the effective fitness landscape at our level of description, where stochastic sub-processes ---like antigen capture from follicular dendritic cells, antigen presentation on the B cell surface, and the encounter and interaction with T-helper cells--- are not resolved \cite{Victora-Nussenzweig}. 

Motivated by these findings, by some evidence of epistasis in the binding affinity of antibodies to antigens \cite{Thomas-bnab-2021,Thomas-bnab-2023, AT-epistasis, compensatory-mut-antibody}, and for the sake of tractability, we study here a maximally epistatic model known as \emph{House of Cards} or Random Energy Model (REM) \cite{HoC-Kingman,Franz-Peliti-REM,Derrida-REM, Derrida-PRB}, where the fitness values associated to each node in the graph are independent random variables drawn from an identical distribution: $\pi_\Gamma(\bold f) = \prod_{i} p_\Gamma(f_i)$.
Under this assumption, 
the asymptotic growth rate distributions in Eqs.~\eqref{rho-deloc}--\eqref{rho-loc} are guaranteed to converge to universal laws as $N\to\infty$. 

Thanks to the generalized central limit theorem, the desired distribution in the delocalized regime reads
\begin{equation}
    \rho^{del}_\Gamma(x)\approx h_{\alpha_\Gamma,\beta_\Gamma}\left(\frac{x-{\mu_{\Gamma,N}}}{\sigma_{\Gamma,N}}\right),
    \quad {\rm for}\,N\gg1 \,,
    \label{CLT}
\end{equation}
where $h_{\alpha,\beta}(z)$ is a stable distribution (for $0<\alpha\leq 2$, $-1\leq \beta\leq 1$),  whose canonical representation is given in terms of its characteristic function \cite{Gnedenko-Kolmogorov}. 
When the parent fitness variable has a finite variance, Eq.~\eqref{CLT} reduces to the standard central limit theorem, where $\alpha=2$ and $\beta$ is irrelevant, and $h_{2,\beta}(z)$ corresponds to the Gaussian distribution.

Similarly, it is known from the theory of extreme value statistics that the c.d.f. of the maximum of a set of IID variables converges (for most distributions) to one of three types of functions, rewritten in compact form as:
\begin{equation}
    \int^x_{-\infty} du\, \rho^{loc}_{\Gamma}(u) \approx G_{\gamma_\Gamma}\left(\frac{x-b_{\Gamma,N}}{a_{\Gamma,N}}\right), \quad {\rm for}\,,N\gg1 \,,
\end{equation}
where 
\begin{equation}
    G_\gamma(z)= \begin{cases}
        e^{-(1+\gamma z)^{-1/\gamma}}\qquad \gamma\neq 0,\ 1+\gamma z \geq 0\\
        e^{-e^{-z}} \qquad\quad\quad\ \ \ \gamma =0
    \end{cases}\,,
    \label{G-gamma}
\end{equation}
and $b_{N,\Gamma}$ and $a_{N,\Gamma}$ are, respectively, average and standard deviation of $\max_{i=1\dots N }\{f_i\}$ \cite{EVS-applications}. 

In summary, in both the localized and delocalized cases, the c.d.f. of the mass growth rate is of the form
\begin{equation}
    P_\Gamma(x) \approx \Phi_{K_\Gamma}\left(\frac{x-\Delta_{\Gamma,N}}{\Sigma_{\Gamma,N}}\right),
    \label{univ}
\end{equation}
where $K_\Gamma$ is a set of parameters indicating the shape and skewness of the asymptotic growth rate distribution in the two considered regimes. Specifically, $K_\Gamma=(\alpha,\beta)$ in the delocalized limit, for the stable distribution $h_{\alpha,\beta}(x)$; $K_\Gamma =\gamma$ in the localized limit, for the extreme value distribution $G'_\gamma(x)$. The two sets of parameters $\Delta_{\Gamma,N}$ and $\Sigma_{\Gamma,N}$ represent respectively the shift and scale parameters of the variable of interest (average or maximum); they will depend on $N$ and on the parameters of the parent fitness distribution.

Using the functional form of Eq.~\eqref{univ} in Eq.~\eqref{Pfix}, we obtain
\begin{widetext}
    \begin{equation}
    P_{fix,\Gamma^*}\approx \int dx \frac{\partial }{\partial x}\left({\Phi_{K_{\Gamma^*}}\left(\frac{x-\Delta_{\Gamma^*,N}}{\Sigma_{\Gamma^*,N}}\right)}^{m_{\Gamma^*}}\right)\prod_{\Gamma\neq\Gamma^*}{\Phi_{K_\Gamma}\left(\frac{x-\Delta_{\Gamma,N}}{\Sigma_{\Gamma,N}}\right)}^{m_\Gamma}, \qquad N\gg 1.
    \label{univ-Pfix}
\end{equation}
\end{widetext}
Equation~\eqref{univ-Pfix} shows that, thanks to universality, the fixation probability of each class depends only on a handful of parameters derived from the fitness distributions $p_\Gamma(f)$, and on the germline abundance of the different BCR classes.
For any combination of these model parameters, we can use Eq.~\eqref{univ-Pfix} to study the inter-class competition in the asymptotic time limit, and identify the class with the highest fixation probability as the dominant one.
We can then immediately construct universal immunodominance phase diagrams in the joint parameter space of the limit distributions associated to the asymptotic growth rates of all the considered classes.

\begin{figure}
    \centering
    \includegraphics[width=\textwidth,trim={0.55cm 0 0.2cm 0.5cm},clip]{  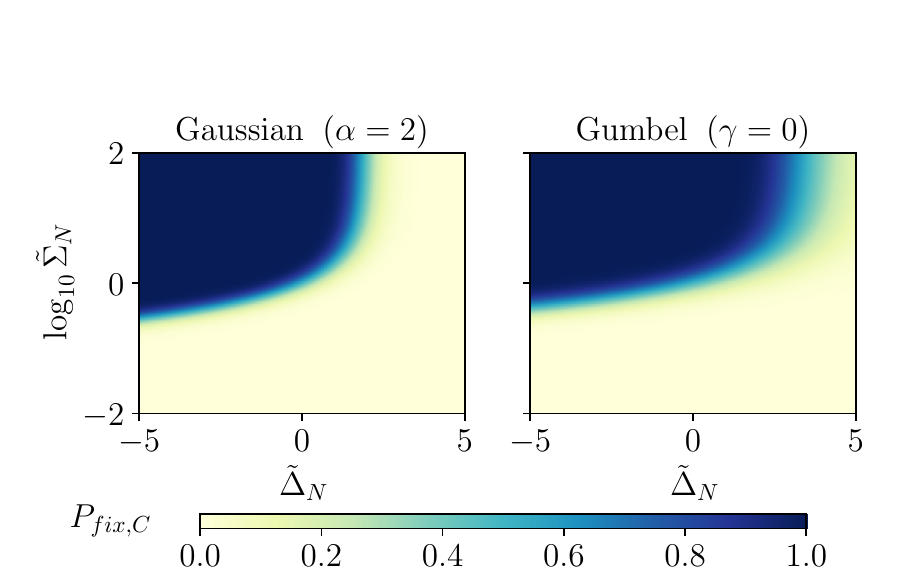}\llap{\parbox[b]{1.6
in}{\small{LOCALIZED}\\\rule{0ex}{1.9in}}}\llap{\parbox[b]{4.7in}{\small{DELOCALIZED}\\\rule{0ex}{1.9in}}}
    \caption{Fixation probability for the two-class competition problem ($C$ vs. $V$), in the case of REM fitness landscapes, with asymptotic mass growth rate distributions of the two classes $C$ and $V$ belonging to the same universality class ($K_C=K_V$ in Eq.~\eqref{univ}). Specifically, we take Gaussian limit stable laws ($\alpha_C=\alpha_V=2$) in the delocalized regime, and Gumbel extreme value distributions ($\gamma_C=\gamma_V=0$) in the localized regime. In this plot, $m_V=m_C=0.5M$, $M=50$.}
    \label{fig:4panel-univ}
\end{figure}

Let us first focus on the simplest, two-class problem, where only two distinct epitopes are considered, targeted by two classes of BCRs named $V$ (variable) and $C$ (conserved).
We are interested in determining when each one of the epitopes is dominant or recessive. 
Plots of $P_{fix,C}$ for some example combinations of universal distributions are shown in Fig.~\ref{fig:4panel-univ} as a function of the following dimensionless combinations of the distribution parameters \footnote{We choose to define $\tilde\Delta_N$ in a non-symmetric way for reasons that will be clear in the study of the parametric 3-class problem, Sec.~\ref{sec:optimization}}:
\begin{equation}
    \tilde \Delta_N = \frac{\Delta_{V,N}-\Delta_{C,N}}{\Sigma_{C,N}}\quad\text{and}\quad
    \tilde \Sigma_N = \frac{\Sigma_{C,N}}{\Sigma_{V,N}}.
\label{params}
\end{equation}
Assuming that the GC capacity $M$ is constrained by the total antigen amount and resource availability, the variables of our problem are only $\tilde\Delta_N$, $\tilde\Sigma_N$, and $m_C/M$. These parameters represent the axes of a three-dimensional immunogenic space, where different \emph{phases} can be identified, as the immunodominant epitope switches from $C$ to $V$ (cf. Fig.~\ref{fig:4panel-univ}). 

Note that the parameters appearing in Eq.~\eqref{univ-Pfix}, or their dimensionless combinations in Eq.~\eqref{params}, all refer to the distributions of the asymptotic growth rates.
The practical question is how to relate these parameters of the universal distributions to the parameters of the parent fitness distributions $p_\Gamma(f)$? 
To answer this question, it is useful to resort to derivations of the stable limit laws and extreme value distributions based on the renormalization group (RG) method. This idea has been repeatedly explored in the literature \cite{Amir_2020,Jona-Cimento,Jona-RG-prob,Sinai-CLT,FS-EVD-PRL,FS-EVD-PRE,EVD-RG-Calvo} and used to derive finite-size corrections to the limit distributions, as well as the asymptotic scaling of shift and scale parameters ---corresponding to $\Delta_{\Gamma,N}$ and $\Sigma_{\Gamma,N}$ in our notation. 
A concise derivation of the flow equations in the space of probability density functions in both the localized and delocalized case is reported in Appendix \ref{app:C}. 

In the delocalized case we restrict to the Gaussian case; other cases refer to parent fitness distributions with diverging first or second moments, which are not biologically relevant.
Identifying $\log N=s$ (treated as a continuous variable for large $N$) and rescaling the sums in such a way that first and second moments do not change with $N$ (see Appendix \ref{app:C}), we find the following flow equations for the parameters of the asymptotic mass growth rate:
\begin{equation}
    \partial_s \Delta_\Gamma(s) = 0,\qquad \partial_s \Sigma_\Gamma(s) = -\frac{1}{2}\Sigma_\Gamma(s).
\end{equation}
Recalling the definition in Eq.~\eqref{params} for the two-class problem, 
\begin{equation}
    \partial_s \tilde\Delta(s)=\frac 1 2\tilde\Delta(s),\qquad \partial_s \tilde\Sigma(s)= 0.
    \label{Gflow}
\end{equation}

In the localized regime, $\Delta_{\Gamma,N}$ and $\Sigma_{\Gamma,N}$ can be computed if we know the cumulative density function
\begin{equation}
    F_\Gamma(z)=\int^z_{-\infty} df p_\Gamma(f) \equiv e^{-e^{-\varphi_\Gamma(z)}},
\end{equation}
via the set of equations (see Appendix~\ref{app:C}, \cite{EVS-applications}):
\begin{equation}
    \label{DS}\Delta_{\Gamma}(s)=\varphi_\Gamma^{-1}(s);\quad  \Sigma_{\Gamma}= \Delta_\Gamma'(s).
\end{equation}
From Eqs.~\eqref{DS}, the flow equations for $\tilde\Delta(s)$ and $\tilde\Sigma(s)$ read
\begin{flalign}
    \label{Dflow}\partial_s \tilde\Delta(s) &= \frac{1}{\tilde \Sigma(s)} -1 - \gamma_C(s) \tilde \Delta(s)\,,\\
    \label{Sflow}\partial_s \tilde\Sigma(s) &= \left[\gamma_C(s) - \gamma_V(s)\right] \tilde\Sigma(s)\,,
\end{flalign}
where $\gamma_\Gamma(s) = \Sigma_\Gamma'(s)/\Sigma_\Gamma(s)\to \gamma_\Gamma$ as $s\to\infty$, for $\Gamma \in \{C,V\}$ . Asymptotically,
\begin{flalign}
    \tilde \Sigma_N &\approx \tilde\Sigma_0 N^{\gamma_C-\gamma_V}\,,\\
    \tilde \Delta_N &\approx \tilde\Delta_0N^{-\gamma_C}-\frac{1-N^{-\gamma_C}}{\gamma_C}+\frac{N^{-\gamma_C}}{\gamma_V\tilde\Sigma_0}\left(N^{\gamma_V}-1\right)\,,
\end{flalign}
where the initial conditions $\tilde\Sigma_0=\Sigma_C(0)/\Sigma_S(0)$ and $\tilde \Delta_0=\left(\Delta_S(0)-\Delta_C(0)\right)/\Sigma_C(0)$ are defined after the following relations:
\begin{equation}
    F_\Gamma(-\Sigma_\Gamma(0)/\Delta_\Gamma(0))=F_\Gamma'(-\Sigma_\Gamma(0)/\Delta_\Gamma(0))=e^{-1}.
\end{equation}

The flow Eqs.~\eqref{Gflow}, \eqref{Dflow} and \eqref{Sflow} can be exploited to construct an immunodominance phase diagram in the space of `bare' control parameters $\tilde\Delta_0$, $\tilde\Sigma_0$ ---which can be directly reconstructed from the parent distribution $p_\Gamma(f)$--- rather than $\tilde\Delta_N$, $\tilde\Sigma_N$. 
The idea is to identify from these equations the stable fixed points and their basins of attraction, and to associate to each basin the asymptotic value of $P_{fix,C}$ computed at the corresponding fixed point. 

\begin{figure*}
    \centering
    \includegraphics[width=\textwidth,trim={1cm 0 1cm 0},clip]{  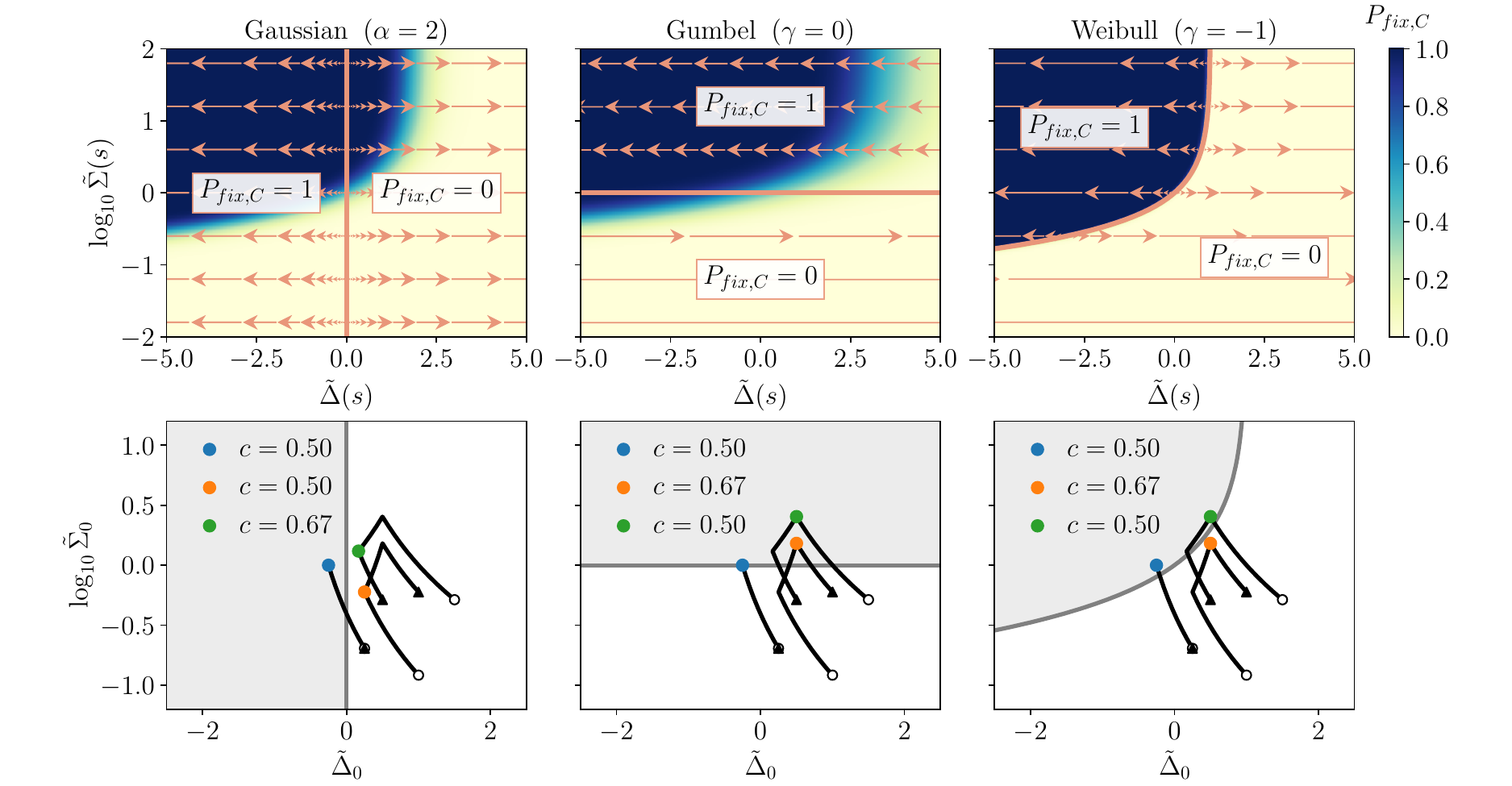}\llap{\parbox[b]{11.15in}{\small{DELOCALIZED}\\\rule{0ex}{3.9in}}}\llap{\parbox[b]{5.in}{\small{LOCALIZED}\\\rule{0ex}{3.9in}}}\llap{\parbox[b]{9.6in}{\small{a}\\\rule{0ex}{3.5in}}}\llap{\parbox[b]{5.45in}{\small{b}\\\rule{0ex}{3.5in}}}\llap{\parbox[b]{1.35in}{\small{c}\\\rule{0ex}{3.5in}}}\llap{\parbox[b]{9.6in}{\small{d}\\\rule{0ex}{1.65in}}}\llap{\parbox[b]{5.45in}{\small{e}\\\rule{0ex}{1.65in}}}\llap{\parbox[b]{1.35in}{\small{f}\\\rule{0ex}{1.65in}}}\llap{\parbox[b]{2.65in}{I\\\rule{0ex}{0.52in}}}\llap{\parbox[b]{1.95in}{II\\\rule{0ex}{0.5in}}}\llap{\parbox[b]{1.55in}{III\\\rule{0ex}{.85in}}}\llap{\parbox[b]{6.78in}{I\\\rule{0ex}{0.52in}}}\llap{\parbox[b]{6.05in}{II\\\rule{0ex}{0.5in}}}\llap{\parbox[b]{5.65in}{III\\\rule{0ex}{.85in}}}\llap{\parbox[b]{10.9in}{I\\\rule{0ex}{0.52in}}}\llap{\parbox[b]{10.2in}{II\\\rule{0ex}{0.5in}}}\llap{\parbox[b]{9.75in}{III\\\rule{0ex}{.85in}}}
    \caption{\textbf{First row}: Identification of the basins of attraction of the RG flow equations for the fixed points $\tilde\Delta=-\infty$ ($P_{fix,C}=1$) and $\tilde\Delta=\infty$ ($P_{fix,C}=0$) in three example cases: \textbf{a} - Delocalized Gaussian case; \textbf{b} - Localized case, with $\gamma_C=\gamma_V=0$; \textbf{c} - Localized case, with $\gamma_C=\gamma_V=-1$. On the background, we show the fixation probability of class $C$ in the space of asymptotic parameters $\tilde \Delta_N,\tilde \Sigma_N$, for $M=50$, $m_C = m_V = 0.5M$. In the foreground, solutions of the flow equations \eqref{Gflow} (a) and \eqref{Dflow}--\eqref{Sflow} (b--c) are shown in orange. 
    \textbf{Second row}: Parametric exploration of the immunogenic space along the lower bound in Eq.~\eqref{bound}. The shaded region indicates the $C$ immunodominant phase in the two-class setting. The black parametric curves are described by Eqs.~\eqref{Deff}--\eqref{Seff}, with parameters: (I) (symmetric case) $\Delta_{A,1}=\Delta_{B,2}=\Sigma_C$; $\Delta_C = 0.75\Sigma_C$; $\Sigma_{A,1}=\Sigma_{B,2}=2\Sigma_C$ ; (II) (asymmetric $\Sigma$) $\Delta_{A,1}=\Delta_{B,2}=1.5\Sigma_C$; $\Delta_C$; $\Sigma_{A,1}=2.5\Sigma_C$; $\Sigma_{B,2}=1.25\Sigma_C$; (III) (asymmetric $\Delta$) $\Delta_{A,1}=\Sigma_C$; $\Delta_{B,2}=2\Sigma_C$; $\Delta_C=0.5\Sigma_C$; $\Sigma_{A,1}=\Sigma_{B,2}=1.33\Sigma_C$. For each curve, the empty circle denotes the $c=0$ point, the filled triangle the $c=1$ point. \textbf{d} - Delocalized regime (Gaussian): here the transition line is vertical and the colored circles indicate the leftmost points along the parametric curves. These points correspond to $c^*=1/2$ when $\Delta_{A,1}=\Delta_{B,2}$ and to $c^*\neq1/2$ when $\Delta_{A,1}\neq\Delta_{B,2}$. \textbf{e} - Localized regime, $\gamma=0$ (Gumbel): here the transition line is horizontal, and thus sensitive to imbalances of the kind $\Sigma_{A,1}\neq\Sigma_{B,2}$. \textbf{f} - Localized regime, $\gamma=-1/2$ (Weibull): colored points represent here the farthest points from the transition line, in the immunodominant phase of epitope $C$.}
    \label{fig:flow}
\end{figure*}

Figure~\ref{fig:flow} depicts the flow equations and of the resulting immunodominance phase diagrams in some illustrative special cases. 
If we work in the delocalized regime and assume, as in Fig.~\ref{fig:4panel-univ}, that the tails of the two parent fitness distributions are of the same type, so that the shape parameters associated to $V$ and $C$ converge to the same value at the same rate, then $\tilde \Sigma_\infty$ is a constant and $\tilde \Delta_\infty$ diverges, keeping the same sign as $\tilde \Delta_0$. Then we obtain only two asymptotic values for $P_{fix,C}$, respectively equal to 0 (for $\tilde\Delta_0>0$) or 1 (for $\tilde\Delta_0<0$). 
In the localized case, when $\gamma_V(s)=\gamma_C(s)\to \gamma$, $\tilde \Sigma$ is a constant and 
\begin{equation}
\tilde \Delta=\frac{1}{\gamma}\left(1/\tilde \Sigma -1\right)
\end{equation}
identifies a line of fixed points, for $\gamma\neq0$, whose stability is determined by the sign of $\gamma$. 
When $\gamma<0$ the points are unstable \footnote{One can reconstruct the phase diagram $P_{fix,C}$ in the plane of control parameters $(\tilde \Delta_0,\tilde \Sigma_0)$ when $\gamma_V(s)=\gamma_C(s)\to \gamma >0$, but it will not exhibit a sharp phase transition. Even in the $N\to\infty$ limit, $P_{fix,C}$ will be a smooth function of $\tilde\Sigma_0$, which must be reconstructed by calculating Eq.~\eqref{univ-Pfix} along the fixed point line.}: since $\lim_{\tilde \Delta\to\infty}P_{fix,C}=0$ and $\lim_{\tilde \Delta\to-\infty}P_{fix,C}=1$, the line of fixed points  becomes a transition line between a phase where the $C$ epitope is dominant and a phase where it is subdominant.

Let us finally note that in the $N\to\infty$ limit the parameter $m_C/M$, indicating the fraction of class $C$ germlines in the GC, becomes irrelevant: the phase diagram is thus the same for all values of $m_C/M\in (0,1)$, as long as $M\ll N$. Finite size corrections may however be important: it is known that the slow convergence to the asymptotic distribution in the size $N$ of the data set is problematic for the quantitative application of extreme value theory to real situations \cite{fisher_tippett_1928}. In our problem, finite size effects may become especially non-negligible if the effective size of the uncorrelated sequence space explored by each clonal lineage is not too big, as it can happen in the case of short-lasting affinity maturation or if we allow for strongly correlated fitness landscapes. 

This analysis, which we illustrated for the two-class problem, can be extended to multiple classes of BCRs targeting an increasing number of distinct epitopes. 

\subsubsection{General principles for optimal antigen cocktail design}\label{sec:optimization}

The immunogenic space construction introduced in the previous Section can be used to investigate under what conditions antigen cocktails can be used to invert the natural immunodominance relations between conserved and highly mutable epitopes.
Under the assumptions outlined in Sec.~\ref{sec:cocktail}, manipulating the relative antigen concentration $c$ in a two-antigen cocktail describes a parametric exploration of the immunogenic space of the three-class problem, as described by the combination rules for the fitness landscapes of Eq.~\eqref{fit-comp} and precursor frequencies of the various BCR classes in Eq.~\eqref{germline-comp}. 

This space is higher-dimensional than the examples of Sec.~\ref{sec:imm-space}, but if we are only interested in the immunodominance of one epitope (e.g. the conserved epitope targeted by B cells of class $C$), we can exploit a lower bound on $P^{3cl}_{fix,C}(c)$ that casts the parametric three-class problem into an effective 2-class problem:
\begin{equation}
P_{fix,C}^{3 cl}(c) \geq P_{fix,C}^{2 cl}\left(\tilde \Delta^{eff}(c), \tilde \Sigma^{eff}(c), m_C(c) \right)\,,
\label{bound}
\end{equation}
where 
\begin{widetext}
\begin{equation}
    P_{fix,C}(c) \approx \int dx \frac{\partial }{\partial x}\left({\Phi_{K_{C}}\left(\frac{x-\Delta_{C,N}}{\Sigma_{C,N}}\right)}^{m_C(c)}\right){\Phi_{K_A}\left(\frac{x-\Delta_{A,N}(c)}{\Sigma_{A,N}(c)}\right)}^{m_A(c)}{\Phi_{K_B}\left(\frac{x-\Delta_{B,N}(c)}{\Sigma_{B,N}(c)}\right)}^{m_B(c)},
    \label{Pfix-cocktail}
\end{equation}
 \end{widetext}
with
\begin{flalign}
    \Delta_{A,N}(c)=c\Delta_N^{A,1}:&\quad \Delta_{B,N}(c)=(1-c)\Delta_N^{B,2};\\
    \Sigma_{A,N}(c)=c\Sigma_N^{A,1};&\quad\Sigma_{B,N}(c)=(1-c)\Sigma_N^{B,2}.
\end{flalign} 
In Eq.~\eqref{bound} $P_{fix,C}^{2 cl}$ is the fixation probability of $C$ against a single effective class $V$, representing the most successful of $A$ and $B$, with effective parameters
\begin{flalign}
    \label{Deff}\tilde \Delta^{eff}_N(c) &= \frac{\max\{c\Delta_N^{A,1},(1-c)\Delta_N^{B,2}\}-\Delta_N^C}{\Sigma_N^C};\\
    \label{Seff}\tilde \Sigma^{eff}_N(c) &= \frac{\Sigma_N^C}{\max\{c\Sigma_N^{A,1},(1-c)\Sigma_N^{B,2}\}}.
\end{flalign}
Let us note that Eqs.~\eqref{bound}--\eqref{Seff} are valid for any $N$, even though the bound may not be equally tight. 

For the sake of simplicity, we work in the space of `bare' parameters $(\tilde \Delta_0,\tilde\Sigma_0)$, where the immunodominance phase diagram exhibits a sharp transition.
From a comparison between the shape of the parametric curves described by Eqs.~\eqref{Deff}--\eqref{Seff} and that of the transition line in the $N\to\infty$ limit, we can gain insight on several questions of interest, i.e.: given a pair of antigens, what is the optimal cocktail formulation that maximizes the production of class $C$ B cells? 
And what are the conditions under which the optimized cocktail can make an epitope which is naturally immunorecessive de facto immunodominant? 

As regards the optimization of the cocktail composition, an obvious solution exists if the system is symmetric under exchange of $c$ and $1-c$, i.e. if  $p_{A,1}(f)=p_{B,2}(f)$ and $\nu_A=\nu_B$. In that case, the optimum corresponds to a balanced mixture of antigens ($c^*=1/2$) for any finite but large $N$, in both  evolutionary regimes (see Appendix~\ref{app:A4}). When the symmetry is broken, the optimal composition will deviate from the even mixture, in order to balance the competition exerted by classes $A$ and $B$ on class $C$. The specific value of $c^*$ is only implicitly determined for finite $N$ in this asymmetric case, but a qualitative analysis can be extracted from the study of the immunodominance phase diagrams in the $N\to\infty$ limit.
When $\Delta_{A,1}\neq \Delta_{B,2}$ (curve III in Fig.~\ref{fig:flow} d--f), the asymmetry stretches the parametric lower bound curve in the horizontal direction: as a result, this asymmetry is best sensed in the delocalized case, where the transition line is perpendicular to that direction. Similarly, the localized Gumbel case exhibits a transition line in the horizontal direction, making the system most sensitive to asymmetries in the $\Sigma$ parameters of classes $A$ and $B$, which stretch the parametric lower bound curve in the vertical direction (curve II in Fig.~\ref{fig:flow} d--f). 

A sufficient condition for the inversion of the immunodominance herarchy is that the parametric curve of Eqs.~\eqref{Deff}--\eqref{Seff} crosses the manifold $P_{fix,C}^{2cl}=1/2$ for some $c\in(0,1)$. Clearly there is a restricted range of combinations of the original parameter values for the three BCR classes such that this crossing can be achieved. 
In the delocalized limit, this condition is tied to the $\Delta_\Gamma$ parameters of the parent single-antigen fitness distributions, while in the localized limit, for $\gamma=0$, it is constrained by the $\Sigma_\Gamma$ parameters. For values of the shape parameter $\gamma<0$, the conditions for inversion look generally more intricate, unless classes $A$ and $B$ have symmetric distributions for antigens 1 and 2. 

These general guidelines are robust to the specific details of the problem, as they only arise from the identification of asymmetries of epitope-specific fitness distributions and of the universality class of the problem. The bulk of the presented results is based on the assumption $N\gg1$ and on long-time asymptotics: we did not investigate in depth finite size or time corrections. However, we remark that finite time effects may be important in the strongly localized limit, where the effective size $N$ of the explored sequence space is dramatically reduced. In such case, since finite size corrections also extend to the third axis of the immunogenic space ---i.e. germline abundance---, cocktail optimization becomes sensitive to possible imbalances in the relative germline abundances, with greater sensitivity exhibited for smaller $N$.

\section{Conclusion}\label{sec:conclusion}

Understanding epitope immunodominance hierarchies is of paramount importance for developing universal vaccines against highly mutable pathogens and for studying pathogen coevolution in immunized hosts or populations. Using a coarse-grained model for affinity maturation and an asymptotic definition of immunodominance, we proposed here a simple framework to characterize immunodominance  hierarchies from the statistical features (or rather \emph{differences} in them) of epitope-specific fitness landscapes of germinal-center B cells. 

We analyzed the paradigmatic case of Random Energy Model landscapes, showing that in this case, as the dimension of the genotypic space explored during affinity maturation increases, the details of the problem become less and less important and a form of universality emerges ---at least in evolutionary regimes that are far from the localization transition. At long times, the impact of the precursor cell abundance vanishes (provided that the number of precursors per class remains sufficiently large for our deterministic approximation) and the average GC population becomes insensitive to the details of the fitness distributions of the various antibody classes. 
The fixation probability of any class, serving as a proxy for the immunogenic advantage of the targeted epitope, ultimately depends on a small, universally defined set of parameters. 

This set of parameters defines what we refer to as \emph{immunogenic space}, where a phase transition occurs between immuno-dominant and immuno-recessive states for a given epitope. We believe that such construction of an immunodominance phase space, which can be extended to other fitness landscape models, can be a useful tool to visualize and characterize pathogen evolution or to determine under what general conditions the manipulation of vaccine compositions may invert the immunodominance relations between epitopes. 
While, to the best of our knowledge, comprehensive datasets enabling a systematic comparison of the binding affinity landscapes of BCRs across various classes are currently lacking, recent advances in epitope mapping and deep mutational scanning techniques indicate the potential for their acquisition and analysis \cite{nature_dataset, cell_dataset, Thomas-bnab-2021}.

In conclusion, we believe that, despite its extreme simplicity, the framework we have proposed presents a promising avenue for a deeper understanding, prediction, and manipulation of immunodominance relations within the context of affinity maturation. To fully realize its potential, an invaluable input would be  high-throughput experimental data that capture without bias the coarse-grained statistical properties of epitope binding affinities.

\begin{acknowledgments}
We warmly thank Arup Chakraborty for suggesting the research problem and engaging in insightful discussions. MK acknowledges support from NSF grant DMR-2218849. 
\end{acknowledgments}

\appendix

\section{First order perturbation theory for the ground state eigenvalue  of the Anderson model}\label{app:A}

For each clonal lineage $\alpha$, we have from Eq.~\eqref{PAM}
\begin{equation}
    H^\alpha_{ij} = -f_i^\alpha\delta_{ij} -\frac{1}{d}\Lambda_{ij},\quad \Lambda_{ij}=-d\delta_{ij}+A_{ij},
\end{equation}
where $A_{ij}$ is the adjacency matrix of the $d$-dimensional hypercube. Without loss of generality, let us take a zero-mean random diagonal and designate by $\sigma$ the scale parameter of the I.I.D. variables, however defined (e.g. standard deviation, when not diverging). Let us define $\tilde f_i^\alpha=f_i^\alpha/\sigma$ a new random fitness variable with the same type of distribution but unit scale. 
The parameter that controls the Anderson localization transition is $d\sigma=\epsilon$, with $\epsilon_c\sim O(1)$. Two limiting regimes can be identified:
\begin{itemize}
    \item $d\sigma=\epsilon\ll 1$ --- delocalized regime. We can rewrite 
    \begin{equation}
        H_{ij}^\alpha = \frac{1}{d}\left(\epsilon H^1_{ij}+H^0_{ij}\right),
    \end{equation}
    where $H^0_{ij}=-\Lambda_{ij}$ and $H^1_{ij}=-\tilde f_i^\alpha\delta_{ij}$. 
    Spectrum and eigenvectors of the unperturbed Hamiltonian are exactly known for the $d$-dimensional hypercube. However, at first order in perturbation theory we are only interested in the unperturbed ground state eigenvalue, $\lambda^0_0=0$, and the associated eigenvector, $\bold v_0^0=\frac{1}{\sqrt N}(1,1,\dots 1)$, which are common to any graph Laplacian. Given the structure of $\bold v_0^0$, the first order correction for the asymptotic growth rate is,
    \begin{equation}
        x=-\lambda_0 \simeq -\frac{1}{d}\left(\lambda_0^0 + \epsilon \lambda_0^1\right) = \frac{1}{N}\sum_{i=1}^N f_i^\alpha,
    \end{equation}
    from which we obtain the distribution $\rho^{del}(x)$ in Eq.~\eqref{rho-deloc}.
    \item $d\sigma=\epsilon\gg 1$ --- localized regime. Let us rewrite:
    \begin{equation}    H_{ij}^\alpha=\sigma\left(H^0_{ij}+\epsilon^{-1}H^1_{ij}\right)
    \end{equation}
    where the unperturbed Hamiltonian $H^0_{ij}=\left(-\tilde f_i^\alpha +\sigma^{-1}\right)\delta_{ij}$ is already diagonal, with ground state localized on the site with the maximum fitness. The perturbation is $H_{ij}^1 = -A_{ij}$. Since the adjacency matrix has null diagonal entries, at first order we have no correction to the ground state eigenvalue, leading to 
    \begin{equation}
        x \simeq  \max\{f_i^\alpha,i=1,\dots,N\} -1.
    \end{equation}
\end{itemize}
Both results hold true even when the random energy landscape is correlated; what changes, in such case, is only how the distributions of the empirical mean and of the maximum are computed.

\section{Renormalization group equations for sum and extreme value statistics of IID variables}\label{app:C}

For completeness, we present in this appendix a concise derivation of the renormalization group (RG) equations for the shift and scale parameters of the stable laws describing the sum and extreme value statistics of a large set of IID variables. The presentation is largely based on Refs.~\cite{Amir_2020,EVS-applications}.

Let us consider a set of $N$ IID variables $x_1,\dots x_N$ with distribution $p_x$, characteristic function $\phi_x$, and cumulative $F_x$. The assumption underlying the  RG construction is that, upon an $N$-dependent linear transformation of the variable of interest (sum or maximum), the p.d.f. of the transformed variable converges to a well-defined limit. Let us define:
\begin{equation}
    \zeta_N = \frac{\sum_{i=1}^N x_i-b_N}{a_N}; \quad \xi_N = \frac{\max_{i=1\dots N} x_i-b_N}{a_N};
\end{equation}
and correspondingly their characteristic and cumulative functions:
\begin{flalign}
    \label{phip}\phi_{\zeta_N}(t) &= \int d\zeta\, p_{\zeta_N}(\zeta)e^{-it\zeta}=\phi_x\left(\frac{t}{a_N}\right)^Ne^{ib_N\frac{t}{a_N}};\\
    \label{Fp}F_{\xi_N}(z) &= \int^z d\xi\, p_{\xi_N}(\xi)=F_x\left(a_N z+b_N\right)^N.
\end{flalign}

Since the maximum and the sum of a set of IID variables can be obtained iteratively, 
we can divide the set of $N$ variables into $p$ groups of equal size $N'=N/p$ and compute the overall sum as the sum over the sums of each group, and the total maximum as the maximum of the maxima of the $p$ groups. The same procedure can be iterated to compute sums and maxima within groups until we reach groups of size 1. Let us then introduce the flowing functions $\phi(t,p)$ and $F(z,p)$, with $p=1+\epsilon$ parametrizing the flow ($\epsilon$ small):
\begin{flalign}
    \label{phipT}\phi(t,p) &= \phi_x\left(\frac{t}{a_p}\right)^pe^{ib_p\frac{t}{a_p}};\\
    \label{FpT}F(z,p) &= F_x\left(a_p z+b_p\right)^p.
\end{flalign}
For convenience, redefine $s=\log p \approx \epsilon$, $g(t,s)=\log\phi(t,e^s)$ and $f(z,s)=-\log[-\log F(z,e^s)]$, so that Eqs.~\eqref{phipT}--\eqref{FpT} become
\begin{flalign}
    \label{gs}g(t,s) &= e^s\left[g_x\left(\frac{t}{a(s)}\right)+i\frac{t}{a(s)}e^{-s}b(s)\right];\\
    \label{fs}f(z,s) &= f_x\left(a(s)z+b(s)\right)-s.
\end{flalign}

The RG equations for the sum and extreme value statistics are obtained by rewriting, respectively, Eqs.~\eqref{gs} or \eqref{fs} as PDEs for $g(t,s)$  or $f(z,s)$ where $g_x$ and $f_x$ do not explicitly appear any more. This can be achieved by an appropriate manipulation of the partial derivatives of $g(t,s)$ and $f(z,s)$ with respect to their arguments. 
The resulting PDEs read
\begin{flalign}
    \label{PDEg}\partial_s g(t,s) &= g(t,s) - t\frac{a'(s)}{a(s)}\partial_tg(t,s) - it\left[\frac{b(s)}{a(s)}-\frac{b'(s)}{a(s)}\right];\\
    \label{PDEf}\partial_s f(z,s) &= \left[\frac{a'(s)}{a(s)}z +\frac{b'(s)}{a(s)}\right]\partial_z f(z,s)-1.
\end{flalign}
Let us notice that, while $f(z,s)\in\R$, $g(t,s)\in\C$, and Eq.~\eqref{PDEg} must be read as a pair of equations for the real and imaginary part of the flowing function $g(t,s)=u(t,s)+iv(t,s)$. The shift and scale parameters, $b(s)$ and $a(s)$, on the contrary, must be real.  

Recall that the assumption behind the RG construction is the existence of an asymptotically stable fixed point for Eqs.~\eqref{PDEg} and \eqref{PDEf}, requiring that the $s$-dependent coefficients converge to constants as $s\to\infty$. Precisely, let us denote these constant limits of the coefficients of Eq.~\eqref{PDEg} as 
\begin{equation}
    \frac{a'(s)}{a(s)}\to \alpha^{-1}, \quad 
    \frac{b(s)}{a(s)}-\frac{b'(s)}{a(s)}\to \beta,
\end{equation}
so that the invariant solution reads
\begin{equation}
    g(t) =\begin{cases}
     C_1 t + i C_2 t + i\beta t \log |t|\ \quad\ \ \mathrm{if}\ \alpha=1\,,\\
    i\frac{\alpha\beta}{\alpha-1}t + |t|^\alpha\left(C_1+iC_2\frac{t}{|t|}\right)\mathrm{if}\ \alpha\neq 1 \,,
    \end{cases}
    \label{hab}
\end{equation}
where $\alpha>0$ and $C_1,C_2\in\R$ are arbitrary constants which will be fixed by the choice of suitable boundary conditions for the PDE \eqref{PDEg} ---interpreted as physically invariant conditions that relate the parent distribution to the asymptotic stable law, as in traditional RG procedure--- and from constraints coming from the support of the parent distribution $p_x(x)$ \cite{Amir_2020}.

From the specification of these boundary conditions, the flow equations for the scale and shift parameters $a(s)$ and $b(s)$ are also derived. Let us focus here on the case where the parent distribution has finite first and second moments (considerations about the scaling with $s$ of $a(s)$ and $b(s)$ in the general case can be found in \cite{Amir_2020}). Let us firstly impose that, for any $s$, the first moment of the distribution reconstructed from $\phi(t,s)=e^{g(t,s)}$ is equal to zero, i.e. that $\partial_tu(t,s)\vert_{t=0}=\partial_tv(t,s)\vert_{t=0}=0$. Using this condition in Eq.~\eqref{PDEg}, we deduce 
\begin{equation}
    \beta(s) = \frac{b(s)}{a(s)}-\frac{b'(s)}{a(s)} =0\ \forall s\ \iff b(s) = b(0)e^s \,,
\end{equation}
where $e^s$ can be identified with the number of variables of which we are computing the sum, and $b(0)$ is the initial shift for the parent distribution, chosen in such a way that $\partial_tv(t,s)\vert_{t=0}=0$. Trying to impose the same condition on Eq.~\eqref{hab}, it is clear that this is only possible for $\alpha>1$. Let us now impose the boundary conditions $\partial^2_{tt} u (t,s)\vert_{t,0}=1$, $\partial^2_{tt} v (t,s)\vert_{t,0}=0$, corresponding to fixing the second moment to be equal to 1. From Eq.~\eqref{PDEg}, this condition implies
\begin{equation}
    \alpha(s)=2 \ \forall s \ \iff a(s) = a(0) (e^s)^{1/2},
\end{equation}
and fixes the values of the arbitrary constants to $C_1=1$, $C_2=0$. 

For the coefficients of Eq.~\eqref{PDEf}, let us denote the limits
\begin{equation}
    \frac{a'(s)}{a(s)}\to \gamma, \quad \frac{b'(s)}{a(s)} \to \delta 
\end{equation}
and the corresponding invariant solution 
\begin{equation}
    f(z)=\frac{1}{\gamma}\log(\gamma z + \delta) + C,
\end{equation}
for $\gamma z+\delta>0$, where again $C$ is an arbitrary constant fixed by the boundary conditions of the PDE \eqref{PDEf}. Let us impose in this case that $f(0,s)=0\ \forall s$ and $\partial_z f(z,s)\vert_{z=0}=1\ \forall s$, corresponding to the conditions $F(0,s)=\partial_z F(z,s)\vert_{z=0}=e^{-1}\ \forall s$. From these conditions (using Eqs.~\eqref{PDEf} and \eqref{fs}) we obtain
\begin{equation}
    a(s) = b'(s), \quad b(s) = f_x^{-1}(s),
\end{equation}
also implying $\delta=1$ and $C=0$.

To obtain the flow equations for the dimensionless parameters defining the axes of the immunogenic phase space, it is sufficient to identify $\Delta_{\Gamma,N}, \Sigma_{\Gamma,N}$ with $b(s), a(s)$ in the localized/extreme value regime, and with $b(s)e^{-s}, a(s)e^{-s}$ in the localized/average regime ($s=\log N$).

\section{Symmetric optimization}\label{app:A4}

When $P_{fix,C}(c)=P_{fix,C}(1-c)$, the symmetry of the problem imposes that $c=1/2$ be a stationary point. That this stationary point is also a maximum and no spontaneous symmetry breaking occurs must be proved by the concavity of this function. 

Let us rewrite Eq.~\eqref{Pfix} for the symmetric case:
\begin{widetext}
\begin{equation}
    P_{fix,C}(c) = m_CM\int dx  \rho_C(x) P_C(x)^{m_CM-1} P_A(x;c)^{m_A(c)M} P_B(x;c)^{m_B(c)M},
\end{equation}
\end{widetext}
where $m_C$ is independent of $c$, and $m_{A}= c\nu m_C$, $m_B = (1-c)\nu m_C$, with $\nu=\nu_{A}/\nu_C=\nu_{B}/\nu_C$. Because of the symmetry of the effective cocktail fitness distributions, 
\begin{equation}
    p_A(f;c) = p_B(f;1-c) \implies  P_A(x;c) = P_B(x;1-c),
\end{equation}
independently of the regime in which affinity maturation occurs. Working in the same setting as in the main text, where $f^A=cf^{A,1}$ and $f^B=(1-c)f^{B,2}$, meaning that replication of a BCR is conditioned to the encounter of the reactive epitope, we can deduce that $x^A/c$ and $x^B/(1-c)$ are equal in probability. Here $x^A$ and $x^B$ represent the asymptotic mass growth rates of hypercubes in class $A$ and $B$. For all the considered limiting regimes, the transformations that map $f$ to $x$ are indeed linear ---extremum for localized regime at asymptotic times; identity for localized regime at early times; average for delocalized regime---, allowing us to rewrite the cumulative distribution of growth rates for classes $A$ and $B$ in terms of a reference distribution for the variable $\hat x = x^A/c$ or $\hat x = x^B/(1-c)$, independent of $c$:
\begin{equation}
    P_A(x;c) = \hat P\left(\frac{x}{c}\right);\qquad P_B(x;c) = \hat P\left(\frac{x}{1-c}\right).
\end{equation}
We can now recognize that $\hat P(x)^{\nu m_C M}\equiv P(x)$ is the cumulative distribution of the maximum of a sequence of $\nu m_C M$ I.I.D. variables distributed as $\hat x$. 
Analogously, $m_CM \rho_C(x) P_C(x)^{m_CM-1}\equiv \tilde \rho(x)$ is the p.d.f. of the maximum of $m_C M$ independent variables distributed as $x^C$. Therefore
\begin{equation}
    P_{fix,C}(c) = \int dx  \tilde\rho(x) P\left(\frac{x}{c}\right)^cP\left(\frac{x}{1-c}\right)^{1-c}.
\label{Pfix-sym}
\end{equation}

It is not guaranteed, for any p.d.f. $\tilde \rho$ and any c.d.f. $P$, that \eqref{Pfix-sym} is a concave function at $c^*=1/2$. When the fixation probability is twice differentiable w.r.t. $c$, the general condition reads $\frac{\partial^2 P_{fix,C}}{\partial c^2}\big\vert_{c^*=\frac{1}{2}}\leq0$, with
\begin{equation}
    \frac{\partial^2 P_{fix,C}}{\partial c^2}\Big\vert_{c^*=\frac{1}{2}} = \int dx \tilde \rho(x) 16x^2P(2x)\frac{\partial^2\log  P(z)}{\partial z^2}\Big\vert_{z=2x}.
\end{equation}
Nonetheless, if we identify $\hat P(a_Nz+b_N)$ with $G_\gamma(z)$ in Eq.~\eqref{G-gamma}, the concavity condition can be easily proved.
In any of the three cases, $\gamma=0$, $\gamma<0$ or $\gamma>0$, the former identification implies that $P(a_{N\nu m_C M}z+b_{N\nu m_C M})$ can also be identified with $G_\gamma(z)$. Hence we obtain, with a simple change of variables:
\begin{widetext}
\begin{equation}
    \frac{\partial^2 P_{fix,C}}{\partial c^2}\Big\vert_{c^*=\frac{1}{2}} = \int dx \tilde \rho_N(x) 16\left(x+\frac{b_N}{2}\right)^2\left(G_\gamma(z)\frac{\partial^2\log  G_\gamma(z)}{\partial z^2}\right)\Big\vert_{z=2x}, \quad \mathrm{with}\quad \tilde\rho_N(x) = \int ds \tilde\rho(s)\delta\left(x-\frac{s-b_N}{a_N}\right).
\end{equation}
We see from Eq.~\eqref{G-gamma} that $G_\gamma(z)$ is not differentiable in $\R$ when $\gamma\neq0$; however we can rewrite in this case 
\begin{equation}
    G_\gamma(z)=\begin{cases}
        e^{-(1+\gamma z)^{-1/\gamma}}\Theta(1+\gamma z),& \gamma >0\\
        e^{-(1+\gamma z)^{-1/\gamma}}\Theta(1+\gamma z) + \Theta(-1-\gamma z), & \gamma <0
    \end{cases}\implies \log G_\gamma(z)=
        -(1+\gamma z)^{-1/\gamma}\Theta(1+\gamma z),\quad \mathrm{for}\ \gamma \neq0\,,
\end{equation}
and approximate the Heaviside functions with smooth sigmoidal functions, e.g. $\Theta_l(s)=\frac{1}{\pi}\tan^{-1}(s/l)+1/2$, such that $\Theta(s)=\lim_{l\to0}\Theta_l(s)$. 
As a result, 

\begin{dmath}
    \frac{\partial^2 P_{fix,C}}{\partial c^2}\Big\vert_{c^*=\frac{1}{2}}=\lim_{l\to0}\begin{cases}
         -\int dx \tilde \rho_{N}(x) 16 \left(x+\frac{b_N}{2}\right)^2 e^{-2x-e^{-2x}}, &\gamma=0;\\
        -\int dx \tilde \rho_N(x)  16 \left(x+\frac{b_N}{2}\right)^2G_\gamma(2x)\left(1+2\gamma x\right)^{-\frac{1}{\gamma}}\bigg[(1+\gamma) \left(1+2\gamma x\right)^{-2} \Theta_l(1+2\gamma x) -\\
        \qquad   2\gamma \left(1+2\gamma x\right)^{-1}\Theta'_l(1+2\gamma x) + \gamma^2 \Theta_l''(1+2\gamma x) \bigg], &\gamma\neq0.
    \end{cases}
    \label{hderiv}
\end{dmath}
\end{widetext}
When $\gamma=0$ $P_{fix,C}$ has a negative well-defined second derivative, which guarantees that $c^*=1/2$ is a smooth maximum. On the contrary, when $\gamma\neq0$, $P_{fix,C}(c)$ can have a kink at $c^*=1/2$. 
By definition, $\lim_{l\to0}\Theta_l(s)=\Theta(s)$, and $\lim_{l\to0}\Theta'_l(s)=\delta(s)$. Hence the first term in the bracket has the sign of $1+\gamma$. This term is finite when $\gamma>0$, while it can diverge for $\gamma<0$ ---depending on whether and how fast $\tilde \rho_N(s)$ converges to 0 for $s\to -1/2\gamma$. The second term is null for any $\gamma>-1$. In the third term, $\Theta_l''$ formally converges to the derivative of a Dirac's delta function, which yields again a null integral for any $\gamma>-1$. In conclusion, $\frac{\partial^2 P_{fix,C}}{\partial c^2}\big\vert_{c^*=\frac{1}{2}}\leq0$ for any $\gamma>-1$, and it possibly diverges for $-1<\gamma<0$. 

One can also make a hand-wavy argument for the concavity of $P_{fix,C}(c)$ in the delocalized regime: in the limit $N\gg 1$, $\hat P(x)$ is a monotonically increasing function that varies very steeply between 0 and 1. Thus we can roughly approximate $\hat P(x)\simeq \theta(x-x^*)$: if $\nu m_C M\ll N$, this implies $P(x)\simeq \theta(x-x^*)$, at the same order of approximation. Therefore
\begin{equation}
    P_{fix,C}(c) \simeq \int_{\max\{cx^*,(1-c)x^*\}}\tilde\rho(x),
\end{equation}
implying $\arg\max_cP_{fix,C}(c) = \arg\min_c\max\{cx^*,(1-c)x^*\}=1/2\ \forall x^*\in \R_0$. The same line of reasoning can be applied to cases where we have more than two BCR classes competing with the cross-reactive one, which target equally immunogenic variants of the variable epitope. In this scenario, the optimum $\bold c^*=\arg\min_{\bold c: \sum_i c_i=1}\max\{c_ix^*\}_{i=1,\dots,E}$ still corresponds to the balanced cocktail.

\end{document}